\documentclass{iopart}
\usepackage{indentfirst}
\usepackage{amsfonts}
\usepackage{amssymb}
\usepackage{bm}
\usepackage{a4wide}
\usepackage{graphicx}
\usepackage{cite}
\usepackage{makeidx}
\usepackage{multicol}
\usepackage{iopams}
\usepackage{bm}

\begin{document}

\title[Method of wave equations exact solutions in matter...]
{Method of wave equations exact solutions in studies of neutrinos and electrons
interaction in dense matter}

\author{A I Studenikin}
\address{Department of Theoretical Physics,
Moscow State University, 119992 Moscow, Russia}
\ead{studenik@srd.sinp.msu.ru}

\begin{abstract}
We present quite a powerful method in investigations of different phenomena that can appear when neutrinos and
electrons propagate in background matter. This method implies use of exact solutions of modified Dirac equations that
contain the correspondent effective potentials accounting for the matter influence on particles. For several particular
cases the exact solutions of modified Dirac and Dirac-Pauli equations for a neutrino and an electron in the background
environment of different composition are obtained (the case of magnetized matter is also considered). Neutrino
reflection, trapping, neutrino pair creation and annihilation in matter and neutrino energy quantization in a rotating
medium are discussed. The neutrino Green functions in matter are also derived. The two recently proposed mechanisms of
electromagnetic radiation by a neutrino and an electron in matter (the spin light of neutrino and electron, $SL\nu$ and
$SLe$) are considered. A possibility to introduce an effective ``matter-induced Lorentz force" acting on a neutrino and
an electron is discussed. A new mechanism of electromagnetic radiation that can be emitted by an electron moving in the
neutrino background with non-zero gradient of density is predicted.

\end{abstract}

\vspace{2pc}

\date{}

\section{Introduction}
The problem of particles interactions under an external
environment influence, provided by the presence of external
electromagnetic fields or media, is one of the important issues of
particle physics. In addition to possibility for better
visualization of fundamental properties of particles and their
interactions being imposed by influence of an external conditions,
the interest to this problem is also stimulated by important
applications to description of different processes in astrophysics
and cosmology, where strong electromagnetic fields and dense
matter may play an important role.

The aim of this paper is to present a rather powerful method in
investigations of different phenomena that can appear when
neutrinos are moving in the background matter
\cite{StuTerPLB05,GriStuTerPLB05_GC05}. In addition, we also
demonstrate how this method can be applied to electrons moving in
background matter
\cite{StuJPA06,StuAFdB06,GriShiStuTerTro_12LomCon,StuNeu_06,GriStuTerTroShiIzvVuz07}.
The developed new approach \cite{StuAFdB06} establishes a basis
for investigation of different phenomena which can arise when
neutrinos and electrons move in dense media, including those
peculiar for astrophysical and cosmological environments.

 The method discussed is based on the use of the modified Dirac equations for
 the particles wave
functions, in which the correspondent effective potentials
accounting for matter influence on the particles are included. It
is similar to the Furry representation \cite{FurPR51} in quantum
electrodynamics, widely used for description of particles
interactions in the presence of external electromagnetic fields.
In this technique, the evolution operator $U_{F}(t_1, t_2)$, which
determines the matrix element of the process, is represented in
the usual form
\begin{equation}
U_{F} (t_1, t_2)= T exp \bigg[-i \int
\limits_{t_1}^{t_2}j_{\mu}(x) A^{\mu}{d}x  \bigg],
\end{equation}
where $A_{\mu}(x)$ is the quantized part of the potential
corresponding to the radiation field, which is accounted within
the perturbation-series techniques. At the same time, the electron
(a charged particle) current is represented in the form
\begin{equation}
j_{\mu}(x)={e \over 2}\big[\overline \Psi_e \gamma _{\mu}, \Psi_e
\big],
\end{equation}
where $\Psi_e$ are the exact solutions of the Dirac equation for
the electron in the presence of external electromagnetic field
given by the classical non-quantized potential $A_{\mu}^{ext}(x)$:
\begin{equation}\label{D_eq_QED}
\left\{ \gamma^{\mu}\big(i\partial_{\mu} -eA_{\mu}^{cl}(x)\big) -
m_e \right\}\Psi_e (x)=0.
\end{equation}

Note that within this approach the interaction of charged
particles with the external electromagnetic field is taken into
account exactly while the radiation field is allowed for by
perturbation-series expansion techniques. A detailed discussion of
the use of this method can be found in \cite{SokTerSynRad68}. Many
processes with electrons under the influence of external
electromagnetic fields were investigated using this method. In
particular, this method was applied \cite{e_dispersion_AMM_B_matt}
for derivation of an electron dispersion relation in external
electromagnetic fields as well as in studies of the problem of the
electron anomalous magnetic moment in external fields (see
\cite{StuPPN90} for a review).

In Section 2.1 we derive the modified Dirac equation for the neutrino wave
function in the presence of matter and find its exact solutions including the
neutrino energy spectrum (Section 2.2). On this basis we discuss the neutrino
reflection, trapping and also neutrino pair annihilation and creation in matter
(Section 2.3). In Section 2.4 we consider the modified Dirac equation for the
case when neutrino propagates in rotating matter and find that its energy is
quantized very much similar to the electron energy Landau quantization in a
magnetic field. In Section 2.5 we consider the Dirac-Pauli equation for a
neutrino moving in matter. The correspondent neutrino energy spectrum, as well
as the one of the modified Dirac equation, can be used for obtaining the
correct values for the flavour and helicity neutrino energy difference in
matter (Section 2.6). In Section 2.7 we use the modified Dirac-Pauli equation
to get neutrino energy spectrum in magnetized and polarized matter. Section 2.8
is devoted to discussion of the modified Dirac equation  for a Majorana
neutrino in matter. Neutrino Green functions, for both Dirac and Majorana
cases, are derived in Section 3. In Section 4 we apply the developed method of
exact solutions of the quantum wave equation to the study of an electron moving
in background matter and found exact solutions of the correspondent Dirac
equation. In Section 5 we illustrate how the obtained exact solutions can be
used in studies of different processes in matter. As two examples, we discuss
evaluation of quantum theory of the spin light of neutrino ($SL\nu$) and spin
light of electron ($SLe$) in matter, the two recently discussed new mechanisms
of electromagnetic radiation produced by a neutrino and an electron moving in
matter. A possibility to introduce an effective ``matter induced Lorentz force"
acting on a neutrino and an electron is discussed in conclusions Section 6. We
also predict a new mechanism of electromagnetic radiation that can be emitted
by an electron moving in the neutrino background with non-zero gradient of
density. The proposed mechanism of the electromagnetic radiation can be
important in physics of neutron stars, gamma-ray bursts and black holes.

\section{Quantum equations for neutrino in matter}
\subsection{Modified Dirac equation for neutrino in matter}

 In \cite{StuTerPLB05}  (see  also
\cite{GriStuTerPLB05_GC05, StuJPA06}) we derived the modified
Dirac equation for neutrino wave function exactly accounting for
the neutrino interaction with matter. Let us consider the case of
matter composed of electrons, neutrons, and protons and also
suppose that the neutrino interaction with background particles is
given by the standard model supplied with the singlet right-handed
neutrino. The corresponding addition to the neutrino effective
interaction Lagrangian is given by
\begin{equation}\label{Lag_f}
\Delta L_{eff}=-f^\mu \Big(\bar \nu \gamma_\mu {1+\gamma^5 \over
2} \nu \Big), \ \ f^\mu={\sqrt2 G_F }\sum\limits_{f=e,p,n}
j^{\mu}_{f}q^{(1)}_{f}+\lambda^{\mu}_{f}q^{(2)}_{f},
\end{equation}
where
\begin{equation}\label{q_f}
\fl q^{(1)}_{f}=
(I_{3L}^{(f)}-2Q^{(f)}\sin^{2}\theta_{W}+\delta_{ef}), \
q^{(2)}_{f}=-(I_{3L}^{(f)}+\delta_{ef}), \  \delta_{ef}=\left\{
\begin{tabular}{l l}
1 & for {\it f=e}, \\
0 & for {\it f=n, p}. \\
\end{tabular}
\right.
\end{equation}
Here $I_{3L}^{(f)}$ and $Q^{(f)}$ are, respectively,  values of
the isospin third components and the electric charges of matter
particles ($f=e,n,p$). The corresponding currents $j_{f}^{\mu}$
and polarization vectors $\lambda_{f}^{\mu}$ are
\begin{equation}\label{j}
j_{f}^\mu=(n_f,n_f{\bf v}_f),
\ \ \ \lambda_f^{\mu} =\Bigg(n_f ({\bm \zeta}_f {\bf v}_f ), n_f
{\bm \zeta}_f \sqrt{1-v_f^2}+ {{n_f {\bf v}_f ({\bm \zeta}_f {\bf
v}_f )} \over {1+\sqrt{1- v_f^2}}}\Bigg),
\end{equation}
where $\theta _{W}$ is the Weinberg angle. In the above formulas
(\ref{j}), $n_f$, ${\bf v}_f$ and ${\bm \zeta}_f \ (0\leq |{\bm
\zeta}_f |^2 \leq 1)$ stand, respectively, for the invariant
number densities, average speeds and polarization vectors of the
matter components. Using the standard model Lagrangian with the
extra term (\ref{Lag_f}), we derive the modified Dirac equation
for the neutrino wave function in matter \cite{StuTerPLB05}:
\begin{equation}\label{new} \Big\{
i\gamma_{\mu}\partial^{\mu}-\frac{1}{2}
\gamma_{\mu}(1+\gamma_{5})f^{\mu}-m \Big\}\Psi(x)=0.
\end{equation}
This is the most general form of the equation for the neutrino
wave function in which the effective potential
$V_{\mu}=\frac{1}{2}(1+\gamma_{5})f_{\mu}$ includes both the
neutral and charged current interactions of neutrino with the
background particles and which can also account for effects  of
matter motion and polarization. It should be mentioned that other
modifications of the Dirac equation were previously used in
\cite{ManPRD88,NotRaf88,NiePRD89,ChaZiaPRD88,PanPLB91-PRD92WeiKiePRD97,
OraSemSmoPLB89,HaxZhaPRD91} for studies of the neutrino dispersion
relations, neutrino mass generation and neutrino oscillations in
the presence of matter.

\subsection{Neutrino quantum states in matter}

In the further discussion below we consider the case when matter
is compose of electrons and no electromagnetic field is present in
the background. We also suppose that the matter is unpolarized,
$\lambda^{\mu}=0$. Therefore, the term describing the neutrino
interaction with the matter is given by
\begin{equation}\label{f}
f^\mu=\frac{\tilde{G}_{F}}{\sqrt2}(n,n{\bf v}),
\end{equation}
where we use the notation $\tilde{G}_{F}={G}_{F}(1+4\sin^2 \theta
_W)$.

For the stationary states of the equation (\ref{new}) we get
\cite{StuTerPLB05}
\begin{equation}\label{stat_states}
\Psi({\bf r},t)=e^{-i(  E_{\varepsilon}t-{\bf p}{\bf r})}u({\bf
p},E_{\varepsilon}),
\end{equation}
where $u({\bf p},E_{\varepsilon})$ is independent on the
coordinates and time. Upon the condition that the equation
(\ref{new}) has a non-trivial solution, we arrive to the energy
spectrum of a neutrino moving in the background matter:
\begin{equation}\label{Energy}
  E_{\varepsilon}=\varepsilon \eta {\sqrt{{\bf p}^{2}\Big(1-s\alpha \frac{m}{p}\Big)^{2}
  +m^2} +\alpha m} ,
\end{equation}
where we use the notation
\begin{equation}\label{alpha}
  \alpha=\frac{1}{2\sqrt{2}}{\tilde G}_{F}\frac{n}{m},
\end{equation}
and also introduce the value
$\eta=$sign$\big(1-s\alpha\frac{m}{p}\big)$ in order to provide a
proper behavior of the wave function in the hypothetical massless
case. The values $s=\pm 1$ specify the two neutrino helicity
states, $\nu_{+}$ and  $\nu_{-}$. In the relativistic limit the
negative-helicity neutrino state is dominated by the left-handed
chiral state ($\nu_{-}\approx \nu_{L}$), whereas the
positive-helicity state is dominated by the right-handed chiral
state ($\nu_{+}\approx \nu_{R}$).The quantity $\varepsilon=\pm 1$
splits the solutions into the two branches that in the limit of
the vanishing matter density, $\alpha\rightarrow 0$, reproduce the
positive- and negative-frequency solutions, respectively. It is
also important to note that the neutrino energy in the background
matter depends on the state of the neutrino longitudinal
polarization, i.e. in the relativistic case the left-handed and
right-handed neutrinos with equal momenta have different energies.

We get the exact solution of the modified Dirac equation in the
form \cite{StuTerPLB05}
\begin{equation}\label{wave_function}
\Psi_{\varepsilon, {\bf p},s}({\bf r},t)= \frac{e^{-i(
E_{\varepsilon}t-{\bf p}{\bf r})}}{2L^{\frac{3}{2}}} \left(
\begin{array}{c}{\sqrt{1+ \frac{m}{ E_{\varepsilon}-\alpha
m}}} \ \sqrt{1+s\frac{p_{3}}{p}}
\\
{s \sqrt{1+ \frac{m}{ E_{\varepsilon}-\alpha m}}} \
\sqrt{1-s\frac{p_{3}}{p}}\ \ e^{i\delta}
\\
{  s\varepsilon\eta\sqrt{1- \frac{m}{ E_{\varepsilon}-\alpha m}}}
\ \sqrt{1+s\frac{p_{3}}{p}}
\\
{\varepsilon\eta\sqrt{1- \frac{m}{ E_{\varepsilon}-\alpha m}}} \ \
\sqrt{1-s\frac{p_{3}}{p}}\ e^{i\delta}
\end{array}\right),
\end{equation}
 where the energy $E_{\varepsilon}$ is given by (\ref{Energy}),
  $L$ is the normalization length,
and $\delta=\arctan{p_2/p_1}$. In the limit of vanishing density
of matter, when $\alpha\rightarrow 0$, the wave function
(\ref{wave_function}) transforms to the vacuum solution of the
Dirac equation.

Let us now consider in some detail  properties of a neutrino
energy spectrum (\ref{Energy}) in the background matter that are
very important for understanding of the mechanism of the neutrino
spin light phenomena. For the fixed magnitude of the neutrino
momentum $p$ there are two values for the ``positive sign"
($\varepsilon =+1$) energies
 \begin{equation}\label{Energy_nu}
\fl  E^{s=+1}={\sqrt{{\bf p}^{2}\Big(1-\alpha \frac{m}{p}\Big)^{2}
  +m^2} +\alpha m}, \ \
E^{s=-1}={\sqrt{{\bf p}^{2}\Big(1+\alpha \frac{m}{p}\Big)^{2}
  +m^2} +\alpha m},
\end{equation}
that determine the positive- and negative-helicity eigenstates,
respectively. The energies in (\ref{Energy_nu}) correspond to the
particle (neutrino) solutions in the background matter. The two
other values for the energy, corresponding to the negative sign
$\varepsilon =-1$, are for the antiparticle solutions. As usual,
by changing the sign of energy, we obtain the values
\begin{equation}\label{Energy_anti_nu}
\fl   {\tilde E}^{s=+1}={\sqrt{{\bf p}^{2}
  \Big(1-\alpha \frac{m}{p}\Big)^{2}
  +m^2} -\alpha m}, \ \ \
  {\tilde E}^{s=-1}={\sqrt{{\bf p}^{2}
  \Big(1+\alpha \frac{m}{p}\Big)^{2}
  +m^2} -\alpha m},
\end{equation}
that correspond to the positive- and negative-helicity
antineutrino states in the matter. The neutrino dispersion
relations in matter exhibits a very fascinating feature (see also
\cite{ChaZiaPRD88,PanPLB91-PRD92WeiKiePRD97}): the neutrino energy
may has a minimum at non-zero momentum.
It may also happen that the neutrino group and phase velocities
are  oppositely directed. The expressions in (\ref{Energy_nu}) and
(\ref{Energy_anti_nu}) would reproduce the neutrino dispersion
relations of \cite{PanPLB91-PRD92WeiKiePRD97}, if the contribution
of the neutral-current interaction to the neutrino potential were
omitted.

In the general case of matter composed of electrons, neutrons and
protons the matter density parameter $\alpha$ for different
neutrino species is
\begin{equation}\label{alpha}
  \alpha_{\nu_e,\nu_\mu,\nu_\tau}=
  \frac{1}{2\sqrt{2}}\frac{G_F}{m}\Big(n_e(4\sin^2 \theta
_W+\varrho)+n_p(1-4\sin^2 \theta _W)-n_n\Big),
\end{equation}
where $\varrho=1$ for the electron neutrino and $\varrho=-1$ for
the muon and tau neutrinos.

Note that on the basis of the obtained energy spectrum
(\ref{Energy}) the neutrino trapping and reflection, the
neutrino-antineutrino pair annihilation and creation in a medium
can be studied
\cite{ChaZiaPRD88,LoePRL90,PanPLB91-PRD92WeiKiePRD97,GriStuTerNANP_PAN06,
KacPLB98KusPosPLB02KoePLB05}.

\subsection{Neutrino reflection, trapping and neutrino-antineutrino pair
annihilation and creation in matter}

Analysis of the obtained energy spectrum (\ref{Energy_nu}),
(\ref{Energy_anti_nu}) enables us to predict some interesting
phenomena that may appear at the interface of the two media with
different densities and, in particular, at the interface between
matter and vacuum.
\begin{figure}[h]\label{Reflaction}
\begin{center}
\includegraphics[width=0.5\textwidth]{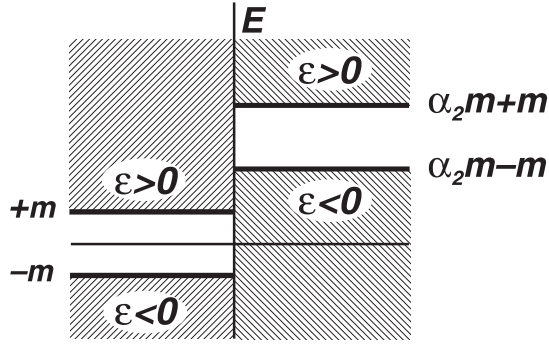}\\
\parbox{0.95\textwidth}{\caption{The interface between the vacuum (left-hand side of the picture)
    and the matter (right-hand side of the picture) with the corresponding
    neutrino band-gaps shown. The parameter $\alpha =\alpha_2 >2$. }}
\end{center}
\end{figure}
Indeed, as it follows from (\ref{Energy_nu}) and
(\ref{Energy_anti_nu}) (see also \cite{GriStuTerNANP_PAN06}), the
band-gap for neutrino and antineutrino in matter is displaced with
respect to the vacuum case in neutrino mass and is determined by
the condition $\alpha m- m\leq E<\alpha m + m$. For instance, if
$\alpha=\alpha_2>2$ then there is  no band-gap overlapping. This
situation is illustrated in Fig.1.

Let us consider first a neutrino moving in the vacuum towards the
interface with energy that falls into the band-gap region in
matter. In this case the neutrino has no chance to survive in the
matter and thus it is reflected from the interface. The same
situation is realized for the antineutrino moving in the matter
with energy falling into the band-gap in the vacuum. In this case
the antineutrino is trapped by the matter. When the energies of
neutrino in the vacuum or antineutrino in the medium fall into the
region between the two band-gaps the effects of the
neutrino-antineutrino annihilation or pair creation may occur (see
the first paper of Ref. \cite{PanPLB91-PRD92WeiKiePRD97} and
\cite{LoePRL90,GriStuTerNANP_PAN06,KacPLB98KusPosPLB02KoePLB05}).
Indeed, the ``negative sign" energy levels in the matter (the
right-hand side of Fig.1) have their counterparts in the
``positive sign" energy levels in the vacuum (the left-hand side
of Fig.1). The neutrino-antineutrino pair creation can be
interpreted as a process a particle state appearance in the
``positive sign" energy range accompanied by appearance of the
hole state in the ``negative sign" energy sea. The phenomenon of
neutrino-antineutrino pair creation in the presence of matter is
similar to the spontaneous electron-positron pair creation the
electrodynamics according (Klein's paradox).

\subsection{Neutrino quantum states in rotating medium}

In this section we apply our method to a particular case when neutrino is
propagating in a rotating medium of constant density \cite{GriSavStuIzvVuz07}.
Suppose that a neutrino is propagating perpendicular to uniformly rotating
matter composed of neutrons. This can be considered for modelling of neutrino
propagation inside a rotating neutron star. The corresponding modified Dirac
equation for the neutrino wave function is given by (\ref{new}) with the matter
potential accounting for rotation,
\begin{equation}\label{rot_f}
f^\mu = -{G}(n,n {\bf v}), \ \ {\bf v}=(\omega y,0,0),
\end{equation}
where $G=\frac{G_F}{\sqrt{2}}$. Here $\omega$ is the angular frequency of
matter rotation around OZ axis, it also is supposed that the neutrino
propagates along OY axis. For  the neutrino wave function components $\Psi (x)
$ we get the from the modified Dirac equation (\ref{new}) a set of
equations{\footnote{The chiral representation for Dirac matrixes is used.},
\begin{equation}
\label{Dir_Comp}
\begin{array}
{rcl} \big[i\left(\partial_0 - \partial_3\right) + G n\big] \Psi_1
+ \big[-\left(i\partial_1 + \partial_2\right) + G n \omega y \big]
\Psi_2
= m \Psi_3, \\
\big[\left(-i\partial_1 + \partial_2 \right) + G n \omega y \big]
\Psi_1 + \big[i\left(\partial_0 +
\partial_3\right) + G n \big] \Psi_2 = m \Psi_4,\\
i\left( \partial_0 + \partial_3 \right) \Psi_3 + \left(i
\partial_1 +
\partial_2 \right) \Psi_4 = m \Psi_1,\\
\left(i \partial_1 - \partial_2\right) \Psi_3 + i \left(\partial_0
-
\partial_3 \right) \Psi_4 = m \Psi_2.\end{array}
\end{equation}
In general case, it is not a trivial task to find solutions of this set of
equations.

The problem is reasonably simplified in the limit of a very small
neutrino mass, i.e. when the neutrino mass can be ignored in the
left-hand side of (\ref{Dir_Comp})  in respect to the kinetic and
interaction terms in the right-hand sides of these equations. In
this case two pairs of the neutrino wave function components
decouple one from each other and four equations (\ref{Dir_Comp})
disintegrate to the two independent sets of two equations, that
couple together the neutrino wave function components in pairs,
($\Psi_1,\ \Psi_2$) and ($\Psi_3,\ \Psi_4$).

The second pair of equations (\ref{Dir_Comp}) does not contain a
matter term and is attributed to the sterile right-handed chiral
neutrino state, $\Psi_{R}$. The corresponding solution can be
taken in the plain-wave form
\begin{equation}
\Psi_{R} \sim L^{-\frac{3}{2}}\exp \{i(- p_0 t + p_1 x + p_2 y +
p_3 z)\}\psi,
\end{equation}
where $p_\mu$ is the neutrino momentum. Then for the components
$\Psi_3$ and $\Psi_4$ we obtain from (\ref{Dir_Comp}) the
following equations
\begin{equation}
\label{PsiR_comp}
\begin{array}{rcl}
\left( p_0 - p_3 \right) \Psi_3 - \left( p_1 - i p_2\right) \Psi_4 = 0, \\

-\left( p_1 + i p_2 \right) \Psi_3 + \left( p_0 + p_3\right)
\Psi_4 = 0.
\end{array}
\end{equation}
Finally, from (\ref{PsiR_comp}) for the sterile right-handed
neutrino we get
\begin{equation}
\label{PsiR} \Psi_R = \frac{\mathrm{e}^{-i p
x}}{L^{3/2}\sqrt{2p_0(p_0 - p_3)}}\left(
\begin{array}{c}
0 \\
0 \\
- p_1 + i p_2 \\
p_3 - p_0
\end{array}
\right),
\end{equation}
where $px=p_{\mu}x^{\mu}, \ p_{\mu}=(p_0,p_1,p_2,p_3)$ and
$x_{\mu} = (t,x,y,z)$. This solution, as it should be, has the
vacuum dispersion relation.

In the neutrino mass vanishing limit the first pair of equations
(\ref{Dir_Comp}) corresponds to the active left-handed neutrino.
The form of these equations is similar to the correspondent
equations for a charged particle (e.g., an electron) moving in a
constant magnetic field $B$ given by the potential $\bm{A} =
(By,0,0)$ (see, for instance, \cite{SokTerSynRad68}). To display
 the analogy, we note that in our case the matter
current components $n\bm{v}$ plays the role of the vector
potential $\bm{A}$. The existed analogy between an electron
dynamics in an external electromagnetic field and a neutrino
dynamics in background matter is further discussed in the
conclusion (Section 6).

The  solution of the first pair of equations (\ref{Dir_Comp}) can
be taken in the form
\begin{equation}
\Psi_{L} \sim \frac{1}{L}\exp \{i(- p_0 t + p_1 x + p_3
z)\}\psi(y),
\end{equation}
and for the components $\Psi_1$ and $\Psi_2$ of the neutrino wave
function we obtain from (\ref{Dir_Comp}) the following equations
\begin{equation}
\label{PsiL_comp}
\begin{array}{rcl}
\Big( p_0 + p_3 + G n \Big) \Psi_1 - \sqrt{\rho}
\left(\frac{\partial }{\partial \eta} - \eta\right) \Psi_2 = 0, \\
\sqrt{\rho}\left(\frac{\partial }{\partial \eta} + \eta\right)
\Psi_1 + \Big(p_0 - p_3 + G n \Big) \Psi_2 = 0,
\end{array}
\end{equation}
where
\begin{eqnarray}
\label{eta} \eta =  \sqrt{\rho} \left(x_2 + \frac{ p_1}{\rho}
\right) , \ \  \rho= G n \omega.
\end{eqnarray}
For the wave function we finally get
\begin{equation} \label{PsiL}
\Psi_L = \frac{\rho^{\frac{1}{4}}\mathrm{e}^{-ip_0 t + ip_1x +
ip_3 z}}{L\sqrt{(p_0 - p_3 + G n)^2+2\rho N}}\left(
\begin{array}{c}
\left(p_0 - p_3 + G n\right) u_N(\eta) \\
- \sqrt{2\rho N} u_{N-1}(\eta) \\
0 \\
0 \end{array} \right),
\end{equation}
where $u_N(\eta)$ are Hermite functions of order $N$. For the
energy of the active left-handed neutrino  we get
\begin{equation}\label{nu_quant_energy}
\label{energy_L} p_0 = \sqrt{p_3^2 + 2 \rho N} - G n, \ \
N=0,1,2,... \ .
\end{equation}
The energy depends on the neutrino momentum component $p_3$ along
the rotation axis of matter and the quantum number $N$ that
determines the magnitude of the neutrino momentum in the
orthogonal plane. For description of antineutrinos one has to
consider the ``negative sign" energy eigeinvalues (see similar
discussion in Section 2.2). Thus, the energy of an electron
antineutrino in the rotating matter composed of neutrons is given
by
\begin{equation}\label{antinu_quant_energy}
\label{energy_L} \tilde p_0 = \sqrt{p_3^2 + 2 \rho N} + G n, \ \ N=0,1,2,... \
.
\end{equation}
Obviously, generalization for different other neutrino flavours and matter
composition is just straightforward (see (\ref{j}) and (\ref{alpha})).

Thus, it is shown \cite{GriSavStuIzvVuz07} that the transversal
motion of an active neutrino and antineutrino is quantized in
moving matter very much like an electron energy is quantized in a
constant magnetic field that corresponds to the relativistic form
of the Landau energy levels (see, for instance, the first book of
\cite{SokTerSynRad68}). Consider again antineutrino. The
transversal motion momentum of is given by
\begin{equation}
\tilde p_{\bot}= \sqrt{2\rho N}.
\end{equation}
The quantum number $N$ determines also the radius of the antineutrino
quasi-classical orbit in matter (it is supposed that $N\gg 1$ and $p_3=0$),
\begin{equation}
R=\sqrt{\frac{2N}{G n \omega}}.
\end{equation}
It follows that antineutrinos can have bound orbits inside a rotating star. To make an estimation of magnitudes, let us
consider a model of a rotating neutron star with radius $R_{NS}=10 \ km$, matter density $n=10^{37} cm^{-3}$ and
angular frequency $\omega=2\pi \times 10^{3} \ s^{-1}$. For this set of parameters, the radius of an antineutrino
orbits is less than the typical star radius $R_{NS}$ if the quantum number $N\leq N_{max} =10^{10}$. Therefore,
antineutrinos that occupy orbits with $N\leq 10^{10}$ can be bounded inside the star. The scale of the bounded
antineutrinos energy estimated by (\ref{antinu_quant_energy}) is of the order $\tilde p_0\sim 1 \ eV$. It should be
underlined that within the quasi-classical approach the neutrino binding on circular orbits is due to an effective
force that is orthogonal to the particle speed. Note that there is another mechanism of neutrinos binding inside a
neutron star when the effect is produced by a gradient of the matter density \cite{LoePRL90}
(see also the conclusion).

\subsection{Modified Dirac-Pauli equation for neutrino in
matter}

To derive the quantum equation for a neutrino wave function in the
background matter we start with the well-known Dirac-Pauli
equation for a neutral fermion with non-zero magnetic moment. For
a massive neutrino moving in an electromagnetic field $F_{\mu
\nu}$ this equation is given by
\begin{equation}\label{D_P}
\Big( i\gamma^{\mu}\partial_{\mu} - m -\frac{\mu}{2}\sigma ^{\mu
\nu}F_{\mu \nu}\Big)\Psi(x)=0,
\end{equation}
where $m$ and $\mu$ are the neutrino mass and magnetic moment
\cite{FujShr80} \footnote{For the recent studies of a massive
neutrino electromagnetic properties, including discussion on the
neutrino magnetic moment, see Ref.\cite{DvoStuPRD04JETP04}},
$\sigma^{\mu \nu}=i/2 \big(\gamma^{\mu }\gamma^{\nu}-\gamma^{\nu}
\gamma^{\mu}\big)$. It worth to be noted here that Eq.({\ref{D_P})
can be obtained in the linear approximation over the
electromagnetic field from the Dirac-Schwinger equation, which in
the case of the neutrino takes the following form
\cite{BorZhuTer88}:
\begin{equation}\label{D_S}
(i\gamma^\mu\partial_\mu-m) \Psi (x) =\ \int M^{F}(x',x)\Psi
(x')dx' ,
\end{equation}
where  $M^{F}(x',x)$ is the neutrino mass operator in the presence
of the external electromagnetic field.

 Recently in a series of our papers
\cite{EgoLobStuPLB00_LobStuPLB01,DvoStuJHEP02} (see also
\cite{StuPAN04_07}) we have developed the quasi-classical approach
to the massive neutrino spin evolution in the presence of external
fields and background matter. In particular, we have shown that
the well known Bargmann-Michel-Telegdi (BMT) equation
\cite{BarMicTelPRL59} of the electrodynamics can be generalized
for the case of a neutrino moving in the background matter and
being also under the influence of external electromagnetic fields.
The proposed new equation for a neutrino, which simultaneously
accounts  for the electromagnetic interaction with external fields
and also for the weak interaction with particles of the background
matter, was obtained from the BMT equation by the following
substitution of the electromagnetic field tensor $F_{\mu\nu}=(\bf
E,\bf B)$:
\begin{equation}\label{sub}
F_{\mu\nu} \rightarrow E_{\mu\nu}= F_{\mu\nu}+G_{\mu\nu},
\end{equation}
where the tensor $G_{\mu\nu}=(-{\bf P},{\bf M})$ accounts for the
neutrino interactions with particles of the environment. The
substitution (\ref{sub}) implies that in the presence of matter
the magnetic $\bf B$ and electric $\bf E$ fields are shifted by
the vectors $\bf M$ and $\bf P$, respectively:
\begin{equation}
\bf B \rightarrow \bf B +\bf M, \ \ \bf E \rightarrow \bf E - \bf
P. \label{11}
\end{equation}
We have also shown \cite{EgoLobStuPLB00_LobStuPLB01, DvoStuJHEP02} how to construct the tensor $G_{\mu \nu}$ with the
use of the neutrino speed, matter speed, and matter polarization 4-vectors.

Now let us consider the case of a neutrino moving in matter
without any electromagnetic field in the background. Starting from
the Dirac-Pauli equation (\ref{D_P}) for a neutrino in
electromagnetic field $F_{\mu \nu}$, we apply the substitution
(\ref{sub}) which now becomes
\begin{equation}\label{sub_1}
  F_{\mu \nu}\rightarrow G_{\mu \nu}.
\end{equation}
As a result of this substitution, we obtain the quantum equation
for the neutrino wave function in the presence of the background
matter in the form
\cite{StuTerQUARKS_04_0410296_GriStuTer_11LomCon}
\begin{equation}\label{D_P_matter}
\Big( i\gamma^{\mu}\partial_{\mu} - m -\frac{\mu}{2}\sigma ^{\mu
\nu}G_{\mu \nu}\Big)\Psi(x)=0,
\end{equation}
that can be regarded as the modified Dirac-Pauli equation.

Consider an explicit solution of the obtained equation
(\ref{D_P_matter}) for the case of an unpolarized matter composed
of only electrons we have
\begin{equation}\label{G_}
G^{\mu \nu}=  \frac{\tilde{G}_{F}}{2\sqrt{2}\mu} \gamma n
\left(\begin{array}{c} {0}\ \ \ \ \ {0} \ \  \ \ \ {0} \  \  \ \ \ {0} \\
\ {0}\ \ \ \ {0}\ \ \ {-\beta_{3}}\ \ \ {\beta_{2}} \\
\ \ {0}\ \ \ \ {\beta_{3}}\ \ \ \ {0}\ \ {-\beta_{1}} \\
{0}\ \ {-\beta_{2}}\ \ \ {\beta_{1}}\ \ \ {0}
\end{array}\right), \gamma=(1-{\bm\beta}^{2})^{-1/2},
\end{equation}
where ${\bm \beta}=(\beta_{1},\beta_{2},\beta_{3})$ is the
neutrino three-dimensional speed and $n$ denotes the number
density of the background electrons. From (\ref{G_}) and two
equations, (\ref{D_P}) and (\ref{D_P_matter}), it is possible to
see that the term $\frac{\tilde{G}_{F}}{2\sqrt{2}\mu} \gamma n{\bm
\beta}$ in (\ref{D_P_matter}) plays the role of the magnetic field
$\bf B$ in (\ref{D_P}). The corresponding neutrino energy spectrum
is
\begin{equation}\label{Energy_1}
  E={\sqrt{{\bf p}^{2}(1+\alpha^{2})+m^2 -2\alpha m p s}},
\ \  \ \   \alpha=
  \frac{1}{2\sqrt{2}}{\tilde G}_{F}\frac{n}{m}.
\end{equation}
This expression can be transformed to the form
\begin{equation}\label{Energy_2}
E=\sqrt{{\bf p}^{2}+m^2\Big(1-s\frac{\alpha p}{m}\Big)^{2}},
\end{equation}
that can be obtained from the neutrino vacuum spectrum by the
formal shift of the neutrino mass
 $ m\rightarrow m\Big(1-s\frac{\alpha p}{m}\Big)$.

The exact solution of the Dirac-Pauli equation  (\ref{D_P_matter})
can be obtained in the following form
\cite{StuTerQUARKS_04_0410296_GriStuTer_11LomCon}:
\begin{equation}\label{wave_function_1}
\Psi_{{\bf p},s}({\bf r},t)=\frac{e^{-i( Et-{\bf p}{\bf
r})}}{2L^{\frac{3}{2}}} \left(\begin{array}{c}{\sqrt{1+
\frac{m-s\alpha p}{E}}} \ \sqrt{1+s\frac{p_{3}}{p}}
\\
{s \sqrt{1+ \frac{m-s\alpha p}{E}}} \ \sqrt{1-s\frac{p_{3}}{p}}\ \
e^{i\delta}
\\
{  s\sqrt{1- \frac{m-s\alpha p}{E}}} \ \sqrt{1+s\frac{p_{3}}{p}}
\\
{\sqrt{1- \frac{m-s\alpha p}{E}}} \ \ \sqrt{1-s\frac{p_{3}}{p}}\
e^{i\delta}
\end{array} \right).
\end{equation}
 In the limit of vanishing matter density, when
$\alpha\rightarrow 0$, this wave function transforms to the vacuum
solution of the Dirac equation.

The obtained neutrino energy spectrum (\ref{Energy_1}), for not
extremely high matter densities $\alpha\frac {pm}{E_{0}^{2}} \ll
1$, yields the correct result for the energy difference $\Delta
E=E(s=-1)-E(s=+1)$ of the two neutrino helicity states:
\begin{equation}\label{delta_Energy}
  \Delta E\approx 2m\alpha \frac{p}{E_{0}},
\end{equation}
where we use the notation $E_0=\sqrt{p^2 +m^2}$. Therefore, on the
basis of the obtained exact solution for the neutrino wave
function in the case of relativistic neutrinos one can derive the
probability of spin oscillations $\nu_{L} \leftrightarrow \nu_{R}$
in transversal magnetic field with the correct form of the matter
term \cite{Akh88LimMar88}.

\subsection{Flavour and helicity neutrino energy difference in
matter}

Although the neutrino energy spectra corespondent to the modified
Dirac and  Dirac-Pauli equations, (\ref{new}) and
(\ref{D_P_matter}), are not the same, an equal result given by
(\ref{delta_Energy}) for the energy difference $\Delta
E=E(s=-1)-E(s=+1)$ of the two neutrino helicity states can be
obtained from both of the spectra in the low matter density or
high energy limit $\alpha\frac {pm}{E_{0}^{2}} \ll 1$.

It should be also noted that for the relativistic neutrinos the
energy spectrum for the neutrino helicity states of
Eq.(\ref{Energy}) in the low density limit  reproduces the correct
energy values for the neutrino left-handed and right-handed chiral
states:
\begin{equation}\label{E_L}
  E_{\nu_L} \approx E(s=-1)\approx E_0 +{{\tilde {G}}_F \over \sqrt{2}}n,
\ \  \  E_{\nu_R} \approx E(s=-1)\approx E_0,
\end{equation}
as it should be for the active left-handed and sterile
right-handed neutrino in matter.

We should like to note, that the obtained spectra for the flavour
neutrinos of different helicities in the presence of matter
enables one to reproduce the well-known result for the energy
difference of two flavour neutrinos in matter. In order to
demonstrate this we expand the expressions for the relativistic
electron and muon neutrino energies (given by (\ref{Energy}) for
the Dirac case or by (\ref{Energy_Majorana}) below for the
Majorana case), over $m/p\ll 1$ and get
\begin{equation}E_{\nu_e, \nu_{\mu}}^{s=-1}\approx E_0
+ 2\alpha_{\nu_e, \nu_{\mu}} m.
\end{equation}
Then the energy difference for the two active flavour neutrinos is
\begin{equation}\Delta E =
E_{\nu_e}^{s=-1}-E_{\nu_{\mu}}^{s=-1} = \sqrt2 G_F n_e.
\end{equation}
Analogously, considering the spin-flavour oscillations
$\nu_{e_L}\rightleftarrows \nu_{\mu_R}$, for the corresponding
energy difference we find:
\begin{equation}\Delta E =
E_{\nu_e}^{s=-1}-E_{\nu_{\mu}}^{s=+1} = \sqrt2 G_F
\big(n_e-{1\over 2}n_n\big).
\end{equation}
These equations enable one to get the expressions for the neutrino
flavour and spin-flavour oscillation probabilities with resonance
dependence on matter density in the complete agreement with the
results of \cite{WolPRD78MikSmiYF85,Akh88LimMar88}.

\subsection{Modified Dirac-Pauli equation in magnetized and
polarized matter}

It is also possible to generalize the Dirac-Pauli equation
(\ref{D_P}) (or (\ref{D_P_matter})) for the case when a neutrino
is moving in a magnetized background matter. For this case (i.e.,
when the effects of matter and magnetic field on neutrino have to
be accounted for simultaneously) the modified Dirac-Pauli equation
is \cite{StuTerQUARKS_04_0410296_GriStuTer_11LomCon}
\begin{equation}\label{D_P_matter_1}
\Big\{ i\gamma^{\mu}\partial_{\mu} - m -\frac{\mu}{2}\sigma ^{\mu
\nu}(F_{\mu \nu}+G_{\mu \nu})\Big\}\Psi(x)=0.
\end{equation}
The neutrino energy in the magnetized matter can be obtained
from (\ref{Energy_1}) by the following redefinition
\begin{equation}\label{redefin}
  \alpha \rightarrow \alpha '=\alpha
  +\frac{\mu B_{\parallel}}{p},
\end{equation}
where $B_{\parallel}= ({\bf B} {\bf p})/ p$ is the longitudinal to
the neutrino momentum magnetic field component. Thus, the neutrino
energy in this case reads
\begin{equation}\label{energy_2}
E=\sqrt{{\bf p}^{2}+m^2\Big(1-s\frac{\alpha p +\mu
B_{\parallel}}{m}\Big)^{2}}.
\end{equation}
For the relativistic neutrinos the expression of
Eq.(\ref{energy_2}) gives, in the linear approximation over the
matter density and the magnetic field strength, the correct value
(see \cite{EgoLobStuPLB00_LobStuPLB01,DvoStuJHEP02,StuPAN04_07})
for the energy difference of the two opposite helicity states in
the magnetized matter:
\begin{equation}\label{Delta}
\Delta_{eff}= {{\tilde {G}}_F \over \sqrt{2}}n +2{\mu
B_{\parallel} \over \gamma}.
\end{equation}

Note that the problem of the neutrino dispersion relation in an
external magnetic field and matter was  also studied previously in
many papers with use of different methods \cite{nu_dispersion_B}.

Now we can consider the neutrino spin oscillations in the presence
of non-moving matter being under the influence of an arbitrary
constant magnetic field ${\bf B}={\bf B}_{\parallel} +{\bf
B}_{\perp}$, here $\bf B_{\perp}$ is the transversal to the
neutrino momentum component of the external field. In the
adiabatic approximation the probability of the oscillations $\nu_L
\leftrightarrow \nu_R$ can be written in the form,
\begin{equation}\label{P_L_R}
\fl P_{\nu_L \rightarrow \nu_R} (x)=\sin^{2} 2\theta_{eff}
\sin^{2}{\pi x \over L_{eff}}, \ \  sin^{2}
2\theta_{eff}={E^2_{eff} \over {E^{2}_{eff}+\Delta^{2}_{eff}}}, \
\ L_{eff}={2\pi \over \sqrt{E^{2}_{eff}+\Delta^{2}_{eff}}},
\end{equation}
where $E_{eff}=2\mu B_{\perp}$ (terms $\sim \gamma^{-1}$ are
omitted  here), and $x$ is the distance traveled by the neutrino.

Let us now shortly discuss the effect of matter polarization.
Consider the case of matter composed of electrons in the presence
of such strong background magnetic field so that the following
condition is valid $B>\frac{p_{F}^{2}}{2e}$, where
$p_{F}=\sqrt{\mu ^2 - m^{2}_{e}}$,  $\mu$ and $m_e$ are,
respectively, the Fermi momentum, chemical potential, and mass of
electrons. Then all of the electrons occupy the lowest Landau
level, therefore the matter is completely polarized in the
direction opposite to the unit vector $\frac{{\bf B}}{B}$. From
the general expression for the tensor $G_{\mu \nu}$ (see the
second paper of \cite{EgoLobStuPLB00_LobStuPLB01}) we get
\cite{StuTerQUARKS_04_0410296_GriStuTer_11LomCon}
\begin{equation}\label{G_1}
\fl
G^{\mu \nu}=  \frac{1}{2\sqrt{2}\mu} \gamma n
\left\{\tilde{G}_{F}\left(\begin{array}{c} {0}\ \ \ \ \ {0} \ \  \ \ \ {0} \  \  \ \ \ {0} \\
\ {0}\ \ \ \ {0}\ \ \ {-\beta_{3}}\ \  {\beta_{2}} \\
\ \ \  {0}\ \ \ \ {\beta_{3}}\ \ \ \ {0}\ \ {-\beta_{1}} \\
{0}\ \ {-\beta_{2}}\ \ \ {\beta_{1}}\ \ \ {0}
\end{array}\right)-{G}_{F}\left(\begin{array}{c}
{0}\ \ {-\beta_{2}}\ \ \ {\beta_{1}}\ \ {0}\\
\ {\beta_2}\ \ \ \ {0}\ \  {-\beta_{0}}\ \ {0} \\
{-\beta_1}\ \ \ {\beta_{0}}\ \ \ \ {0}\ \ {0} \\
\ \ \ {0}\ \ \ \ \ {0} \ \  \ \  {0} \  \  \  {0}
\end{array}\right)\right\},
\end{equation}
Thus, the modified Dirac-Pauli equation (\ref{D_P_matter_1}) with
the tensor $G_{\mu \nu}$ given by (\ref{G_1}) can be used for
description of the neutrino motion in matter which is magnetized
and totally polarized in respect to the magnetic field vector $\bm
B$ direction. The neutrino energy in such a case can be obtained
from (\ref{Energy_1}) by the following redefinition
\begin{equation}\label{redefin}
  \alpha \rightarrow {\tilde \alpha} =\alpha
  \left[1- \frac{sign \left(\frac{B_{\parallel}}{B}\right)}
  {1+\sin^{2}
  4\theta_{W}}\right]
  +\frac{\mu B_{\parallel}}{p}.
\end{equation}
In Eq.(\ref{redefin}), the second term in brackets  accounts for
the effect of the matter polarization. It follows, that the effect
of the matter polarization can reasonably change the total matter
contribution to the neutrino energy (\ref{energy_2}) (see also
\cite{NunSemSmiValNPB97}).

Note that the problem of the neutrino dispersion relation in an
external magnetic field and matter was previously also studied in
many papers with use of different methods \cite{nu_dispersion_B}.

\subsection{Majorana neutrino}

We have considered so far the case of the Dirac neutrino. Now let
us turn to the Majorana neutrino \cite{GriStuTerNANP_PAN06}. For a
Majorana neutrino we derive the following contribution to the
effective Lagrangian accounting for the interaction with the
background medium
\begin{equation}\label{Lag_f_Majorana}
\Delta L_{eff}=-f^\mu (\bar \nu \gamma_\mu \gamma^5 \nu ),
\end{equation}
which leads to the Dirac equation
\begin{equation}\label{Dirac_Majorana} \Big\{
i\gamma_{\mu}\partial^{\mu}-\gamma_{\mu}\gamma_{5}f^{\mu}-m
\Big\}\Psi(x)=0.
\end{equation}
This equation differs from the one, obtained in the Dirac case, by
doubling of the interaction term and lack of the vector part. The
corresponding energy spectrum for the equation
(\ref{Dirac_Majorana}) is:
\begin{equation}\label{Energy_Majorana}
  E_{\varepsilon}=\varepsilon{\sqrt{{\bf p}^{2}\Big(1-2s\alpha \frac{m}{p}\Big)^{2}
  +m^2}}.
\end{equation}
From this expression it is clear, that the energy of the Majorana
neutrino has its minimal value equal to the neutrino mass, $E =m$.
This means that no effects are anticipated for the Majorana
neutrino such as the Dirac neutrino has at the two media interface
and which are discussed above. So that, in particular, there is no
Majorana neutrino trapping and reflection by matter. It should be
noted that the equation (\ref{Dirac_Majorana}) and the Majorana
neutrino spectrum in matter were discussed previously also in
\cite{PanPLB91-PRD92WeiKiePRD97,
BerVysBerSmiPLB87_89GiuKimLeeLamPRD92GiuKimLeeLamPRD92BerRosPLB94}.

\section{Neutrino Green function in matter}

The neutrino Green function, along with the wave function, is an
important characteristic of the neutrino (propagation) in matter.
Developing further the method of the exact solutions for the
studies of the neutrino propagation in matter, we consider
explicit Green functions for the the modified Dirac equation for
the Dirac and Majorana neutrinos \cite{PivStuPOS05}. For the Dirac
and Majorana neutrino Green functions we obtain the same equations
as for the correspondent wave functions, (\ref{new}) and
(\ref{Dirac_Majorana}), with the only difference that $-\delta(x)$
functions stay on the right hand sides. In the momentum
representation the equation for the Green function has the
following form:
\begin{equation}
\label{Green_f} \Big\{ i\gamma_{\mu}\partial^{\mu}-\frac{1}{2}
\gamma_{\mu}(a+\gamma_{5})f^{\mu}-m \Big\}G(p)=-1 \ \ ,
\end{equation}
where $a=1$ for the Dirac neutrino and $a=0$ for the Majorana
case. Squaring the left hand side operator, it is possible to
obtain the following expression for the Green function of neutrino
in matter:
\begin{equation}
\label{Green_f_final} \fl
G(q)=\frac{-\left(q^{2}-m^{2}\right)\left(\hat q+m\right)+\hat
f\left(\hat q-m\right)P_{L}\left(\hat q+m\right)-f^{2}\hat
q\,P_{L}+2\left(fq\right)P_{R}\left(\hat
q+m\right)}{\left(q^{2}-m^{2}\right)^{2}-2\left(fq\right)
\left(q^{2}-m^{2}\right)+f^{2}q^{2}} \ \ ,
\end{equation}
where
\begin{equation}
\fl q=p-\frac{1}{2}(a-1)f,\ \hat q = q_\mu \gamma^\mu, \
q^2=q_{\mu}q^{\mu},\ \hat f= f_\mu \gamma^\mu, \
(fq)=f_{\mu}q^{\mu},\  P_{L,R}=\frac{1}{2}(1\pm\gamma_{5})\  .
\end{equation}

Now let us consider the denominator of expression
(\ref{Green_f_final}). The poles of the Green function determine
the neutrino dispersion relation. Equating the denominator to
zero, we obtain quadratic equation relative to $q_{0}$:
\begin{equation}
\label{denom_zero}
\left(q^{2}-m^{2}\right)^{2}-2\left(fq\right)\left(q^{2}-m^{2}\right)+f^{2}q^{2}=0.
\end{equation}
In some special cases equation (\ref{denom_zero}) can be solved
analytically. One of such cases is that of uniform medium, moving
at constant speed $\bf{v}$ parallel to the neutrino momentum
$\bf{p}$. In this case we can solve equation (\ref{denom_zero})
for $q_{0}$ and then find $p_{0}$,
\begin{equation}
\label{q0_solution}
{p}_{0}=\frac{1}{2}\left[af_{0}+s\,|\,{\bf{f}}\,|+\epsilon\sqrt{4m^2+\Big(2|\,{\bf{p}}-
\frac{1}{2}(a-1){\bf{f}}\,|-s(f_{0}+s|\,{\bf{f}}\,|)\Big)^{2}}\right].
\end{equation}
There are four solutions of (\ref{q0_solution}) corespondent to
$s=\pm 1$ and $\epsilon =\pm 1$.  From Eq.(\ref{q0_solution}) one
can find, that all solutions except one are of definite sign for
any $|{\bf{p}}|$. The sign of $p_{0}$ for $\epsilon=-1$ and $s=1$
however can be both positive and negative for different
$|{\bf{p}}|$. One can also note, that in case of
\begin{equation}\label{condition}
af_{0}+|\,{\bf{f}}\,|<2m,
\end{equation}
 the sign of this $p_{0}$ is always negative.

 In case condition
Eq.(\ref{condition}) holds, the solution of equations (\ref{new})
and (\ref{Dirac_Majorana}) can be expressed in the form of the
superposition of plane waves each with definite sign of energy.
Note that if the condition (\ref{condition}) is violated then
there exists a plane wave that has positive energy for some
$|{\bf{p}}|$ and negative for others. Stated in other words, the
condition (\ref{condition}) means that Green function
(\ref{Green_f_final}) can be chosen causal by imposing special
rules of poles bypassing (negative poles should be bypassed from
below and positive poles should be bypassed from above). Once we
got the causal Green function the perturbation technique can be
developed for the description of the neutrino propagation in
matter.

Another way to interpret the condition (\ref{condition}) is to
turn attention to \cite{GriStuTerNANP_PAN06} where it was shown,
that for the matter at rest, the spontaneous $\nu\tilde{\nu}$ pair
creation can take place only when $f_{0}>2m$. From the analysis of
the allowed energy zones for neutrino in matter it follows that
$\nu\tilde{\nu}$ pair creation in moving matter can take place
only when $af_{0}+|{\bf{f}}|>2m$. So that the possibility of using
the neutrino Green function (\ref{Green_f_final}) is limited by
the particular value of matter density when $\nu\tilde{\nu}$ pair
creation processes become available.

For the Majorana neutrino moving through uniform matter at rest
the condition (\ref{condition}) is always valid for any matter
densities $f_{0}$ because $a=0$ and ${\bf{f}}=0$ in this case.

\section{Electron wave function and energy spectrum in matter}
In
\cite{StuJPA06,GriShiStuTerTro_12LomCon,StuNeu_06,GriStuTerTroShiIzvVuz07},
it has been shown how the approach, developed at first for
description of a neutrino motion in the background matter, can be
spread for the case of an electron propagating in matter.

Let us consider an electron having the standard model interactions
with particles of electrically neutral matter composed of
neutrons, electrons and protons. This can be used for modelling a
real situation existed, for instance, when electrons move in
different astrophysical environments. We suppose that there is a
macroscopic amount of the background particles in the scale of an
electron de Broglie wave length. To further simplify the model, we
consider the case of nuclear matter
\cite{KacPLB98KusPosPLB02KoePLB05, Bethe71} composed of neutrons.
Then the addition to the electron effective interaction Lagrangian
is
\begin{equation}\label{Lag_f_e}
\Delta L^{(e)}_{eff}=-{f}^\mu \Big(\bar e \gamma_\mu
{1-4\sin^{2}\theta_{W}+\gamma^5 \over 2} e \Big),
\end{equation}
where the explicit form of $f^{\mu}$ depends on the background
particles density, speed and polarization and is determined by
(\ref{Lag_f}) and (\ref{q_f}). The modified Dirac equation for the
electron wave function in matter is \cite{StuJPA06}
\begin{equation}\label{new_e}
\Big\{ i\gamma_{\mu}\partial^{\mu}-\frac{1}{2}
\gamma_{\mu}(1-4\sin^{2}\theta_{W}+\gamma_{5}){\widetilde
f}^{\mu}-m_e \Big\}\Psi_{e}(x)=0,
\end{equation}
where for the case of an electron moving in the background of
neutrons
\begin{equation}
{\tilde f}^{\mu}=-f^\mu=\frac{G_F}{\sqrt
2}(j^{\mu}_n-\lambda^{\mu}_n).
\end{equation}
We consider below unpolarized neutrons so that
\begin{equation}\label{f2_e}
\widetilde{f}^{\mu}=\frac{G_{F}}{\sqrt{2}}(n_n,n_n{\bf v}),
\end{equation}
here $n_n$ is the neutrons number density and $\mathbf v$ is the
speed of the reference frame in which the mean momentum of the
neutrons is zero. The corresponding electron energy spectrum in
the case of unpolarized matter at rest is given by
\begin{equation}\label{Energy_e}
  E_{\varepsilon}^{(e)}=
  \varepsilon \eta_e \sqrt{{{\bf p}}^{2}\Big(1-s_e\alpha_n
  \frac{m_e}{p}\Big)^{2}
  +{m_e}^2} +c{\alpha}_n m_e, \ \ \alpha_n=\frac{1}{2\sqrt{2}}
  {G_F}\frac{n_n}{m_e},
\end{equation}
where $c=1-4\sin^2 \theta_W$,
$\eta_e=$sign$\big(1-s\alpha_n\frac{m_e}{p}\big)$, $m_e$ and $p$
are the electron mass and momentum.

For the wave function of the electron moving in nuclear matter we
get \cite{
GriShiStuTerTro_12LomCon,StuNeu_06,GriStuTerTroShiIzvVuz07}
\begin{equation}
\Psi_{\varepsilon, {\bf p},s}({\bf r},t)=\frac{e^{-i(
E^{(e)}_{\varepsilon}t-{\bf p}{\bf r})}}{2L^{\frac{3}{2}}}
\left(%
\begin{array}{c}{\sqrt{1+ \frac{m_e}{ E^{(e)}_{\varepsilon}-c\alpha_n m_e}}}
\ \sqrt{1+s\frac{p_{3}}{p}}
\\
{s \sqrt{1+ \frac{m_e}{ E^{(e)}_{\varepsilon}-c\alpha_n m_e}}} \
\sqrt{1-s\frac{p_{3}}{p}}\ \ e^{i\delta}
\\
{  s\varepsilon\eta_e\sqrt{1- \frac{m_e}{
E^{(e)}_{\varepsilon}-c\alpha_n m_e}}} \ \sqrt{1+s\frac{p_{3}}{p}}
\\
{\varepsilon\eta_e\sqrt{1- \frac{m_e}{
E^{(e)}_{\varepsilon}-c\alpha_n m_e}}} \ \
\sqrt{1-s\frac{p_{3}}{p}}\ e^{i\delta}
\end{array}%
\right).
\end{equation}
The exact solutions of this equation open a new method for
investigation of different quantum processes which can appear when
electrons propagate in matter.

\section{Neutrino and electron spin light in matter}

In this section we illustrate how the developed method based on
the use of the exact solutions of the modified Dirac equations for
neutrino and electron wave functions can be used in the study of
different phenomena that arris when a neutrino or electron move in
matter. As an example, we discuss below the spin light of neutrino
($SL\nu$) and the spin light of electron ($SL e$), new types of
electromagnetic radiation that can be produced by the Dirac
particles while moving in the background matter. The spin light of
neutrino in matter, one of the four new phenomena studied in our
recent papers (see for a short review \cite{StuNPB05}), is an
electromagnetic radiation that can be emitted by a massive
neutrino (due to its non-zero magnetic moment) when the particle
moves in the background matter. Within the quasi-classical
treatment the existence of the $SL\nu$ was first proposed and
studied in \cite{LobStuPLB03LobStuPLB04DvoGriStuIJMP05} , while
the quantum theory of this phenomenon was developed in
\cite{StuTerPLB05, GriStuTerPLB05_GC05, GriStuTerNANP_PAN06,
StuTerQUARKS_04_0410296_GriStuTer_11LomCon, StuPAN04_07,
StuJPA06,LobPLB05}. The spin light of electron in matter
\cite{StuJPA06,
GriShiStuTerTro_12LomCon,GriStuTerTroShiIzvVuz07,StuNeu_06} also
originates from the particle magnetic moment procession in matter.
Note that the term ``spin light of electron" was used first in
\cite{ITernSPU95} for designation of the particular spin-dependent
contribution to the electron synchrotron radiation power. It
should be stressed that the $SL\nu$ and $SLe$ in matter are really
new mechanisms of electromagnetic radiation of quite a different
nature that ones considered before including the Cherenkov
radiation of particles in medium. In particular, the spin light
processes may proceed even when the photon refractive index in
matter equals to $n_{\gamma}\ = \ 1$ \ .

The corresponding Feynman diagram of these processes is shown in
Fig.2. \begin{figure}[h]\label{diagram}
\begin{center}
{  \includegraphics[scale=.45]{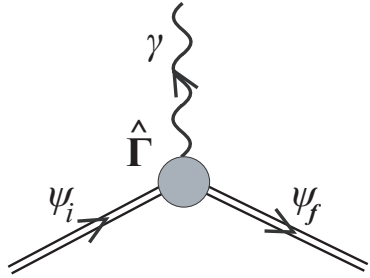}}
    \caption{
    The $SL\nu$ and $SLe$ radiation diagram. }
  \end{center}
\end{figure}
The particles initial $\psi_{i}$ and final $\psi_{f}$ states
(shown by ``broad lines") are exact solutions of the corresponding
Dirac equations for the neutrino and electron in matter that
account for the particles interaction with matter. The amplitude
of the $SL\nu$ process is given by

\begin{equation}\label{amplitude}
\fl
\begin{array}{c} \displaystyle
  S_{f i}=-\mu \sqrt{4\pi}\int d^{4} x {\bar \psi}_{f}(x)
  ({\hat {\bf \Gamma}}{\bf e}^{*})\frac{e^{ikx}}{\sqrt{2\omega L^{3}}}
   \psi_{i}(x), \ \ \ \
     \hat {\bf \Gamma}=i\omega\big\{\big[{\bf \Sigma} \times
  {\bm \varkappa}\big]+i\gamma^{5}{\bf \Sigma}\big\},
\end{array}
\end{equation}
where $\mu$ is the neutrino magnetic moment, $k^{\mu}=(\omega,{\bf
k})$ and ${\bf e}^{*}$ are the photon momentum and polarization
vectors, ${\bm \varkappa}={\bf k}/{\omega}$ is the unit vector
pointing in the possible direction of the emitted photon
propagation. The amplitude of the process $SLe$ is given by
\begin{equation}\label{amplitude_e}
   S_{f i}=-ie \sqrt{4\pi}\int d^{4} x {\bar \psi}_{f}(x)
  ({\gamma}^{\mu}{e}^{*}_{\mu})\frac{e^{ikx}}{\sqrt{2\omega L^{3}}}
   \psi_{i}(x),
\end{equation}
where $-e$ is the electron charge. The further evaluation of the
$SL\nu$ and $SLe$ characteristics of the processes, such as the
differential and total rates and powers, angular distributions
etc, can be found in the mentioned above papers. From the energy
momentum conservation in the $SL\nu$ and $SLe$ processes we obtain
the following values for the spin light radiation energy
\begin{equation}\label{omega}
\fl   \omega_{SL}=\frac{2\alpha_n m_l p \ [\tilde{E}- (p+\alpha_n
m_l)\cos\theta_{SL}]}
   {(\tilde{E}-p\cos\theta_{SL})^2-(\alpha_n m_l)^2}, \ \ \
     \tilde{E}={E}-c\alpha_n m_l, \ \ \alpha_n=\frac{1}{2\sqrt{2}}
  {G_F}\frac{n_n}{m_l},
\end{equation}
where $\theta_{SL}$ is the angle between possible directions of
the radiation and the initial particle momentum ${\bf p}$, for the
case of neutrinos $m_l=m_\nu$ and $c_l=c_{\nu}=1$, whereas for
electrons $m_l=m_e$ and $c_l=c_e=1-4\sin^{2}\theta_{W}$. From
(\ref{omega}) it follows that for the relativistic particles and a
wide range of matter densities (that can be found in diverse
astrophysical and cosmological environments) the energy range of
the $SL\nu$ and $SLe$ may even extend up to energies peculiar to
the spectrum of gamma-rays (see also \cite{StuTerPLB05,StuJPA06}).

For the rate of the $SL\nu$ in the case of ultra-relativistic
neutrinos ($p\gg m$) we
obtained\cite{StuTerPLB05,GriStuTerPLB05_GC05}
\begin{equation}\label{gamma_nu}
\Gamma_{SL\nu} = 4 \mu ^2 \alpha ^2 m_\nu^2 p,  \ \ \ \ \ \ \ \
  {m_\nu}/{p} \ll \alpha \ll {p}/{m_\nu},
\end{equation}
where the matter density parameter $\alpha$ is given by
(\ref{alpha}), in case of negative $\alpha$ the $SL\nu$ can be
emitted by antineutrino. The main properties of the $SL\nu$
investigated in \cite{LobStuPLB03LobStuPLB04DvoGriStuIJMP05,
StuTerPLB05,GriStuTerPLB05_GC05} can be summarized as follows
\cite{GriLobStuTerQuarks06}:
 1) a neutrino with nonzero mass and magnetic moment when
moving in dense matter can emit spin light; 2) in general,
$SL\nu$ in matter is due to the dependence of the neutrino
dispersion relation in matter on the neutrino helicity; 3) the
$SL\nu$ radiation rate and  power depend on the neutrino magnetic
moment and energy, and also on the matter density; 4)  the matter
density parameter $\alpha$, that depends on the type of neutrino
and matter composition,  can be negative; therefore the types of
initial and final neutrino (and antineutrino) states, conversion
between which can effectively produce the $SL\nu$ radiation, are
determined by the matter composition; 5) the $SL\nu$ in matter
leads to the neutrino-spin polarization effect; depending on the
type of the initial neutrino (or antineutrino) and matter
composition the negative-helicity relativistic neutrino (the
left-handed neutrino $\nu_{L}$) is converted to the
positive-helicity neutrino (the right-handed neutrino $\nu_{R}$)
or vice versa; 6) the obtained expressions for the $SL\nu$
radiation rate and power exhibit non-trivial dependence on the
density of matter and on the initial neutrino energy; the $SL\nu$
radiation rate and power are proportional to the neutrino magnetic
moment squared which is, in general, a small value and also on the
neutrino energy, that is why the radiation discussed can be
effectively produced only in the case of ultra-relativistic
neutrinos; 7) for a wide range of matter densities the radiation
is beamed along the neutrino momentum, however the actual shape of
the radiation spatial distribution may vary from projector-like to
cap-like, depending on the neutrino momentum-to-mass ratio and the
matter density; 8) in a wide range of matter densities the $SL\nu$
radiation is characterized by total circular polarization; 9) the
emitted photon energy is also essentially dependent on the
neutrino energy and matter density; in particular, in the most
interesting for possible astrophysical and cosmology applications
case of ultra-high energy neutrinos, the average energy of the
$SL\nu$ photons is one third of the neutrino momentum. Considering
the listed above properties of the $SL\nu$ in matter, we argue
that this radiation can be produced by high-energy neutrinos
propagating in different astrophysical and cosmological
environments.

A remark on the possibility for Majorana neutrino to emit the spin
light in matter should be made. Obviously, due to the absence of
the magnetic moment, such radiation is not expected in this case.
However, considering the transition between two neutrinos of
different flavour, it is possible to produce an analogous effect
via the transition magnetic moment, that  Majorana neutrinos can
posses.

Performing the detailed study of the $SLe$ in neutron matter
\cite{GriStuTerTroShiRPJ07} we have found for the total rate
\begin{equation}\label{gamma_e}
\Gamma_{SLe}=e^2 m_e^2/(2p)\left[\ln\big({4\alpha_n p}/{m_e}\big)-
    {3}/{2}\right],  \ \ \ \ {m_e}/{p}\ll\alpha_n\ll {p}/{m_e},
\end{equation}
where it is supposed that $\ln \frac{4\alpha _n p}{m_e} \gg 1$. It was also found that the relativistic electrons can
loose nearly the whole of its initial energy due to the $SLe$ mechanism.

It should be noted that discussion on possible impact of the background plasma on the $SL\nu$ radiation mechanism have
been started in \cite{GriStuTerPLB05_GC05}. Then effects of plasma for the $SL\nu$ and $SLe$ were considered by another
authors in \cite{KuzMikhMPL06IJMPA07}. These authors, after we explained in \cite{GriLobStuTerQuarks06} that their
initial conclusion \cite{KuzMikhhp0605114} that in presence of matter the process ``$\nu_{L}\rightarrow
\nu_{R}+\gamma^{\ast}$ is kinematically forbidden" was wrong, obtained the $SL\nu$ rate with account for the photon
dispersion in plasma. In the case of ultra-high energy neutrino (i.e., in the only case when the time scale of the
process can be much less than the age of the Universe) the $SL\nu$ rate of \cite{KuzMikhMPL06IJMPA07} exactly
reproduces our result (\ref{gamma_nu}) obtained in \cite{StuTerPLB05,GriStuTerPLB05_GC05}. The final result for the
$SLe$ total rate in the second paper of \cite{KuzMikhMPL06IJMPA07} in the leading logarithmic term confirms our result
(\ref{gamma_e}) obtained in \cite{GriStuTerTroShiRPJ07}.

\section{Conclusion}

We have considered a framework for treating different interactions of neutrinos
and electrons in the presence of matter. The method developed is based on use
of exact solutions of modified Dirac equations that include correspondent
matter potentials. It has been demonstrated how this method work in
consideration of different quantum processes that can proceed in presence of
matter.

Finally, let us consider the established in Sections 2.4 and 2.5 analogy
between particles dynamics in the presence of electromagnetic fields and
dynamics in matter. The developed semiclassical approach to description of the
matter effect, driven by (electro)weak forces, is valid as long as interactions
of particles with the background is coherent. This condition is satisfied when
a macroscopic amount of the background particles are confined within the scale
of a neutrino or electron de Broglie wave length. For the relativistic
neutrinos or electrons the following condition should be satisfied the
following condition must be satisfied $\label{d_Br} \frac {n}{\gamma_l
m^3_l}\gg 1$, where $n$ is the number density of matter,
$\gamma_l=\frac{E_l}{m_l}$ and $(l=\nu$ or $e )$. In case of varying density of
the background matter, there is an additional condition for applicability of
the approach developed (see, for instance, \cite{LoePRL90,
DvoStuJHEP02,SilShuPP00}). The variation scale of matter density should be much
larger than the de Broglie wavelength, $\label{d_Br_2} |\frac {\bm \nabla
n}{np}|\ll 1$.

We can further develop the established in Section 2.5 analogy between a
neutrino motion in a rotating matter and an electron motion in a magnetic
field. It is possible to explain the neutrino quasiclassical circular orbits as
a result of action of the
 attractive central force,
\begin{equation}
{\bf F}_m^{(\nu)}=q^{(\nu)}_m {\bm \beta} \times {\bf B}_m, \ {\bf B}_m= {\bm
\nabla} \times {\bf A}_m, \ {\bf A}_m=n {\bf v},
\end{equation}
where the effective neutrino ``charge" in matter (composed of neutrons in the
discussed case) is $q^{(\nu)}_m=-G$, whereas ${\bf B}_m$ and ${\bf A}_m$ play
the roles of effective ``magnetic" field and the correspondent ``vector
potential". Like the magnetic part of the Lorentz force, ${\bf F}_m^{(\nu)}$ is
orthogonal to the neutrino speed ${\bm \beta}$.

 It is
possible to generalize the discussed above description of the
matter effect on neutrinos for the case when the matter density
$n$ is not constant. For the most general case the ``matter
induced Lorentz force" is given by
\begin{equation}\label{m_Lor_force}
{\bf F}_m^{(\nu)}=q^{(\nu)}_m {\bf E}_m + q^{(\nu)}_m {\bm \beta} \times {\bf
B}_m,
\end{equation}
where the effective ``electric" and  ``magnetic" fields are
respectively,
\begin{equation}\label{m_E}
{\bf E}_m=-{\bm \nabla}n -{\bf v}\frac{\partial n}{\partial t} -
n\frac{\partial {\bf v}}{\partial t} ,
\end{equation}
and
\begin{equation}\label{m_B}
{\bf B}_m=n {\bm \nabla} \times {\bf v}-{\bf v} \times {\bm
\nabla}n.
\end{equation}
Using (\ref{Lag_f}) and (\ref{q_f}) (see also (\ref{alpha}) ) these expressions
can be generalized for a background composed of different matter species. The
force acting on a neutrino, produced by the first term of the effective
``electric" field in the neutron matter, was considered in \cite{LoePRL90}.
Note that the same quasiclassical treatment of a neutrino motion in the
electron plasma was considered in \cite{SilShuPP00}.

To conclude, we ague that it is also possible to introduce the ``matter induced
Lorentz force" for an electron moving in background matter. The weak forces
acting on a neutrino and an electron in matter are identical. Therefore,
similar to the case of neutrino, we can write for the force acting on an
electron ${\bf F}_m^{(e)}$ in background matter
\begin{equation}\label{m_Lor_force_e}
{\bf F}_m^{(e)}=q^{(e)}_m {\bf E}_m + q^{(e)}_m {\bm \beta} \times {\bf B}_m,
\end{equation}
where appropriate magnitude for the effective electron ``charge" in matter $q^{(e)}_m$ should be used. As it follows
from (\ref{m_E}) and (\ref{m_Lor_force_e}), an accelerating force acts on an electron when it moves in background
matter with nonvanishing gradient of density. Using this observation,  we should like to discuss a new mechanism of
electromagnetic radiation by an electron moving in the neutrino background ($m=\nu$) with non-zero gradient of its
density. This situation can be realized in different astrophysical and cosmology settings. For instance, this
phenomenon can exist when an electron propagates in the radial direction from a compact star object inside a dense
environment composed predominantly of neutrinos that also move in radial direction after they were emitted from a
central part of the star. In this case that total power of the radiation (in the quasiclassical limit) is given by
\begin{equation}\label{I_elctron_nu}
I=\frac{2}{3}q^{(e)}_{\nu}\Big[\frac{{\bf a}^{2}}{(1-\beta^{2})^2} + \frac{({\bf a}{\bm
\beta})^{2}}{(1-\beta^{2})^3}\Big],
\end{equation}
where ${\bm \beta}$ is the electron speed and ${\bf a}$ is the electron acceleration induced by the gradient of the
neutrino background density. We expect that the proposed mechanism of the electromagnetic radiation can be important in
other astrophysics settings like one that can be realized in neutron stars, gamma-ray bursts and black holes.

\section*{Acknowledgments }
 I would like to thank Michael Bordag for
the invitation to participate to the Workshop on Quantum Field Theory under the Influence of External Conditions and
for the kind hospitality provided me in Leipzig. I am also thankful to Veniamin Berezinsky, Gennady Chizhov, Alexander
Dolgov, Carlo Giunti, Alexander Grigoriev, Vladimir Lipunov, Lev Okun, Vladimir Ritus, Valery Rubakov, Olga Ryazhskaya,
Mikhail Vysotsky and Alexei Ternov for useful discussions.

\end{document}